\documentclass[prd,twocolumn,showpacs,preprintnumbers,amsmath,amssymb,superscriptaddress]{revtex4}
\usepackage{graphicx}
\usepackage{amsmath}
\usepackage{amssymb}
\usepackage{afterpage}
\usepackage{graphics}
\usepackage{float}
\usepackage{rotating}
\usepackage{xspace}
\usepackage{dcolumn}%align table columns on decimal point
\usepackage{bm} %bold math
\begin{document}
\preprint{FERMILAB-PUB-08-212-E, BNL-XXXXX-2008-YY, arXiv:ZZZZZZZ [hep-ex]}

\title{Testing Lorentz Invariance and CPT Conservation with NuMI Neutrinos in the MINOS Near Detector}         % Enter your title between curly braces

\newcommand{\Cambridge}{Cavendish Laboratory, University of Cambridge, Madingley Road, Cambridge CB3 0HE, United Kingdom}
\newcommand{\FNAL}{Fermi National Accelerator Laboratory, Batavia, Illinois 60510, USA}
\newcommand{\RAL}{Rutherford Appleton Laboratory, Chilton, Didcot, Oxfordshire, OX11 0QX, United Kingdom}
\newcommand{\UCL}{Department of Physics and Astronomy, University College London, Gower Street, London WC1E 6BT, United Kingdom}
\newcommand{\Caltech}{Lauritsen Laboratory, California Institute of Technology, Pasadena, California 91125, USA}
\newcommand{\ANL}{Argonne National Laboratory, Argonne, Illinois 60439, USA}
\newcommand{\Athens}{Department of Physics, University of Athens, GR-15771 Athens, Greece}
\newcommand{\NTUAthens}{Department of Physics, National Tech. University of Athens, GR-15780 Athens, Greece}
\newcommand{\Benedictine}{Physics Department, Benedictine University, Lisle, Illinois 60532, USA}
\newcommand{\BNL}{Brookhaven National Laboratory, Upton, New York 11973, USA}
\newcommand{\CdF}{APC -- Universit\'{e} Paris 7 Denis Diderot, 10, rue Alice Domon et L\'{e}onie Duquet, F-75205 Paris Cedex 13, France}
\newcommand{\Cleveland}{Cleveland Clinic, Cleveland, Ohio 44195, USA}
\newcommand{\Delhi}{Department of Physics and Astrophysics, University of Delhi, Delhi 110007, India}
\newcommand{\GEHealth}{GE Healthcare, Florence South Carolina 29501, USA}
\newcommand{\Harvard}{Department of Physics, Harvard University, Cambridge, Massachusetts 02138, USA}
\newcommand{\HolyCross}{Holy Cross College, Notre Dame, Indiana 46556, USA}
\newcommand{\IIT}{Physics Division, Illinois Institute of Technology, Chicago, Illinois 60616, USA}
\newcommand{\Indiana}{Indiana University, Bloomington, Indiana 47405, USA}
\newcommand{\ITEP}{High Energy Experimental Physics Department, Institute of Theoretical and Experimental Physics, 
  B. Cheremushkinskaya, 25, 117218 Moscow, Russia}
\newcommand{\JMU}{Physics Department, James Madison University, Harrisonburg, Virginia 22807, USA}
\newcommand{\LASL}{Nuclear Nonproliferation Division, Threat Reduction Directorate, Los Alamos National Laboratory, Los Alamos, New Mexico 87545, USA}
\newcommand{\Lebedev}{Nuclear Physics Department, Lebedev Physical Institute, Leninsky Prospect 53, 119991 Moscow, Russia}
\newcommand{\LLL}{Lawrence Livermore National Laboratory, Livermore, California 94550, USA}
\newcommand{\MIT}{Lincoln Laboratory, Massachusetts Institute of Technology, Lexington, Massachusetts 02420, USA}
\newcommand{\Minnesota}{University of Minnesota, Minneapolis, Minnesota 55455, USA}
\newcommand{\Crookston}{Math, Science and Technology Department, University of Minnesota -- Crookston, Crookston, Minnesota 56716, USA}
\newcommand{\Duluth}{Department of Physics, University of Minnesota -- Duluth, Duluth, Minnesota 55812, USA}
\newcommand{\Oxford}{Subdepartment of Particle Physics, University of Oxford, Oxford OX1 3RH, United Kingdom}
\newcommand{\Pittsburgh}{Department of Physics and Astronomy, University of Pittsburgh, Pittsburgh, Pennsylvania 15260, USA}
\newcommand{\IHEP}{Institute for High Energy Physics, Protvino, Moscow Region RU-140284, Russia}
\newcommand{\RoyalH}{Physics Department, Royal Holloway, University of London, Egham, Surrey, TW20 0EX, United Kingdom}
\newcommand{\Carolina}{Department of Physics and Astronomy, University of South Carolina, Columbia, South Carolina 29208, USA}
\newcommand{\SLAC}{Stanford Linear Accelerator Center, Stanford, California 94309, USA}
\newcommand{\Stanford}{Department of Physics, Stanford University, Stanford, California 94305, USA}
\newcommand{\StJohnFisher}{Physics Department, St. John Fisher College, Rochester, New York 14618 USA}
\newcommand{\Sussex}{Department of Physics and Astronomy, University of Sussex, Falmer, Brighton BN1 9QH, United Kingdom}
\newcommand{\TexasAM}{Physics Department, Texas A\&M University, College Station, Texas 77843, USA}
\newcommand{\Texas}{Department of Physics, University of Texas at Austin, 1 University Station C1600, Austin, Texas 78712, USA}
\newcommand{\TechX}{Tech-X Corporation, Boulder, Colorado 80303, USA}
\newcommand{\Tufts}{Physics Department, Tufts University, Medford, Massachusetts 02155, USA}
\newcommand{\UNICAMP}{Universidade Estadual de Campinas, IF-UNICAMP, CP 6165, 13083-970, Campinas, SP, Brazil}
\newcommand{\USP}{Instituto de F\'{i}sica, Universidade de S\~{a}o Paulo,  CP 66318, 05315-970, S\~{a}o Paulo, SP, Brazil}
\newcommand{\Warsaw}{Department of Physics, Warsaw University, Ho$\dot{z}$a 69, PL-00-681 Warsaw, Poland}
\newcommand{\Washington}{Physics Department, Western Washington University, Bellingham, Washington 98225, USA}
\newcommand{\WandM}{Department of Physics, College of William \& Mary, Williamsburg, Virginia 23187, USA}
\newcommand{\Wisconsin}{Physics Department, University of Wisconsin, Madison, Wisconsin 53706, USA}
\newcommand{\deceased}{Deceased.}

\affiliation{\ANL}
\affiliation{\Athens}
\affiliation{\Benedictine}
\affiliation{\BNL}
\affiliation{\Caltech}
\affiliation{\Cambridge}
\affiliation{\UNICAMP}
\affiliation{\CdF}
\affiliation{\FNAL}
\affiliation{\Harvard}
\affiliation{\IIT}
\affiliation{\Indiana}
%\affiliation{\IHEP}
%\affiliation{\ITEP}
\affiliation{\JMU}
%\affiliation{\Lebedev}
%\affiliation{\LLL}
\affiliation{\UCL}
\affiliation{\Minnesota}
\affiliation{\Duluth}
\affiliation{\Oxford}
\affiliation{\Pittsburgh}
\affiliation{\RAL}
\affiliation{\USP}
\affiliation{\Carolina}
\affiliation{\Stanford}
\affiliation{\Sussex}
\affiliation{\TexasAM}
\affiliation{\Texas}
\affiliation{\Tufts}
\affiliation{\Warsaw}
\affiliation{\Washington}
\affiliation{\WandM}
%\affiliation{\Wisconsin}

\author{P.~Adamson}
\affiliation{\FNAL}
%\affiliation{\UCL}
%\affiliation{\Sussex}

\author{C.~Andreopoulos}
\affiliation{\RAL}
%\affiliation{\Athens}

\author{K.~E.~Arms}
\affiliation{\Minnesota}

\author{R.~Armstrong}
\affiliation{\Indiana}

\author{D.~J.~Auty}
\affiliation{\Sussex}

%\author{S.~Avvakumov}
%\affiliation{\Stanford}

\author{D.~S.~Ayres}
\affiliation{\ANL}

\author{B.~Baller}
\affiliation{\FNAL}

%\author{B.~Barish}
%\affiliation{\Caltech}

%\author{P.~D.~Barnes~Jr.}
%\affiliation{\LLL}

\author{G.~Barr}
\affiliation{\Oxford}

\author{W.~L.~Barrett}
\affiliation{\Washington}

%\author{E.~Beall}
%\altaffiliation[Now at\ ]{\Cleveland .}
%\affiliation{\ANL}
%\affiliation{\Minnesota}

\author{B.~R.~Becker}
\affiliation{\Minnesota}

% ...

\collaboration{The MINOS Collaboration}
\noaffiliation

\date{\today}          % Enter your date or \today between curly braces

\begin{abstract}
A search for a sidereal modulation in the MINOS near detector neutrino data was performed.  If present, this signature could be a consequence of Lorentz and CPT violation as predicted by a class of extensions to the Standard Model.  No evidence for a sidereal signal in the data set was found, implying that there is no significant change in neutrino propagation that depends on the direction of the neutrino beam in a sun-centered inertial frame. Upper limits on the magnitudes of the Lorentz and CPT violating terms in these extensions to the Standard Model lie between 0.01-1\% of the maximum expected, assuming a suppression of these signatures by factor of $10^{-17}$.  
\end{abstract}

\maketitle

At experimentally accessible energies, signals for Lorentz and CPT violation can be described by a class of extensions to the Standard Model often referred to as the SME~\cite{others,CKcomb}.  This SME theory is an observer-independent framework that contains all the Lorentz violating (LV) terms involving particle fields in the Standard Model of particle physics and gravitational fields in the General Theory of Relativity.  Since the Standard Model is thought to be the low-energy limit of a more fundamental theory that unifies quantum physics and gravity at the Planck scale, $m_P \simeq 10^{19}$ GeV, it is suggested in~\cite{CKcomb} that the violations of Lorentz and CPT symmetries predicted by the SME provide a link to Planck scale physics.  Although the magnitude of LV signatures in the accessible energy limit are suppressed by a factor of order the electroweak scale divided by the Planck scale, $m_W/m_P \sim 10^{-17}$~\cite{KM2}, these low-energy probes of new physics have been explored in many ways with current experimental technologies~\cite{lit}.  

The SME framework predicts several unconventional phenomena, among which is one that arises from the dependence of the neutrino oscillation probability on the direction of neutrino propagation~\cite{KM,KM2}.  For experiments like MINOS~\cite{Adamson:2007gu} with both beam neutrino source and detector fixed on the Earth's surface, the Earth's sidereal rotation causes the direction of neutrino propagation $\hat p$ to change with respect to the Sun-centered inertial frame in which the SME is formulated~\cite{LSND}.  The theory predicts that this rotation introduces a sidereal variation in the number of neutrinos detected from the beam.  The LSND collaboration~\cite{LSND} did not see this signal.  In this paper we use a sample of neutrinos identified in the MINOS near detector (ND) in a search for this sidereal signal.  The neutrinos  were generated by the Neutrinos at the Main Injector (NuMI) neutrino beam at Fermilab~\cite{minos2}, a beam whose flavor composition is 98.7\% $\nu_\mu + \bar \nu_\mu$~\cite{Adamson:2007gu}.  

According to the SME, the probability that one of these $\nu_\mu$ oscillates to flavor $\nu_x$, where $x$ is $e$ or $\tau$, over a distance $L$ from its production to detection due to Lorentz and CPT violation is given by~\cite{KM} 
\begin{eqnarray}
\label{eq:osc}
P_{\nu_\mu \rightarrow \nu_x} &\simeq& L^2 [( {\mathcal C})_{\mu x} +  
({\mathcal A_c})_{\mu x} \cos{(\omega_\oplus T_\oplus)} \nonumber \\ 
& & + 
({\mathcal A_s})_{\mu x}   \sin{(\omega_\oplus T_\oplus)} + 
({\mathcal B_c})_{\mu x}   \cos{(2 \omega_\oplus T_\oplus)}  \nonumber \\
& & + ({\mathcal B_s})_{\mu x}  \sin{(2 \omega_\oplus T_\oplus)}]^2,
\end{eqnarray}
where $\omega_\oplus = 2\pi/(23^h \, 56^m \, 04.0982^s)$ is the Earth's sidereal frequency and $T_\oplus$ is the local sidereal time of the neutrino detection.  For the MINOS ND, $<L> \sim 750$ m.  In this equation, the expressions for $({\mathcal A_c})_{\mu x} $, $({\mathcal A_s})_{\mu x} $, and $({\mathcal C})_{\mu x} $ include both CPT and Lorentz violating terms; the expressions for $({\mathcal B_c})_{\mu x} $ and $({\mathcal B_s})_{\mu x} $ include only Lorentz violating terms.   Since CPT violation implies Lorentz violation in field theory~\cite{G1}, there are no terms that depend on CPT violation alone.  These parameters are combinations of the SME coefficients $(a_L)^\mu$ and $(c_L)^{\mu \nu}$ that describe LV~\cite{KM}.  The magnitude of the coefficients in eq.~(\ref{eq:osc}) also depend on the neutrino propagation direction.  For the NuMI beam line, the direction vectors of the beam are defined by colatitude $\chi = (90^\circ -$ latitude) $= 90.^\circ - 41.84056333^\circ $, $\theta = 93.346^\circ$, and $\phi = 23.909^\circ$.  Here $\theta$ and $\phi$ are the beam zenith and azimuthal angles; $\theta$ is measured from the $z$-axis which points toward the zenith;  $\phi$ is defined from the $x$-axis which is along the detector axis and increases away from the NuMI target.   The $y$-axis makes a right handed coordinate system.  The explicit relationships between the SME coefficients and the beam direction are found in~\cite{KM}.  In eq.(\ref{eq:osc}), the CPT violating terms depend on $L$ and the Lorentz violating terms depend on $L \times E_{\nu}$, where $E_{\nu}$ is the neutrino energy. This unconventional behavior is to be compared with the $L / E_{\nu}$ dependence of the oscillation probability resulting from non-zero neutrino masses~\cite{Nunokawa:2005nx}. 

The MINOS ND~\cite{minosD} is a magnetized 0.98 kton steel scintillator tracking calorimeter that lies 103 m underground at Fermilab and it is  made of  282  $4 \times6$~m$^2$ octagonal planes.  Each plane is comprised of a 2.54 cm thick steel plate and a layer of scintillator strips with dimensions  of 4.1 cm $\times$ 1~cm.  Each scintillator strip is coupled via wavelength-shifting fiber to one pixel of a 64-pixel Hamamatsu M64 PMT~\cite{Tagg:2004bu}.  The ND readout continuously integrates the PMT charges, while the data acquisition accepts data above a threshold of 0.25 photo-electrons.  

Protons with mean energy of 120 GeV$/c$ are extracted from the Main Injector giving a beam spill of 10~$\mu$s duration and period of approximately 2.4~s.  The number of protons delivered to the target for each spill (POT) was measured using toroidal beam current transformers.  The uncertainty in the number of POT for each spill is $\pm 1.0$\%~\cite{Adamson:2007gu}.  Neutrino events in the ND fiducial volume were selected based on their timing and spatial information~\cite{minos2,minosNC}.  The neutrinos were separated into charged current-like (CC) or neutral current-like (NC) events, as described in~\cite{minos2}.  CC $\nu_\mu$ events, identified by a $\mu^{-}$ track~\cite{Adamson:2007gu}, were selected for further analysis to maximize the $\nu_\mu \rightarrow \nu_x$ oscillation signal in MINOS.  

Standard beam quality cuts~\cite{Adamson:2007gu} were applied to select spills for the analysis.  In addition, data quality cuts were applied to remove runs in which there were detector problems, including cooling system failures, magnetic coil failures, or an incorrectly configured readout trigger.

The data were taken during two run periods.  The parameters for these two runs are given in Table~\ref{table:runParam}.  The numbers of events and POT given are the numbers remaining in the sample after all cuts have been made. 

\begin{table}
\caption{\label{table:runParam} Run Parameters}
\begin{tabular}{|c|c|c|c|}

\hline \hline

  & ~~CC Events~~& ~POT~ & ~~Run Dates~~ \\  \hline 

Run I &  $1.82 \times 10^6$  & $1.25 \times 10^{20}$ & May05 -- Feb06\\ 

Run II  & $1.62 \times 10^6$ & $1.14 \times 10^{20}$  & Sept06 -- Mar07  \\

\hline \hline

\end{tabular}

\end{table}

Since the sidereal phase histograms in this analysis require accurate event timing, we describe how time stamps are generated.  The spill time is  determined by the Global Positioning System (GPS) receiver located in the ND hall that reads out absolute Universal Coordinated Time (UTC) and is accurate to 200~ns~\cite{minosToF}.  The Main Injector accelerates protons to 120 GeV$/c$ and the spills are extracted to NuMI using a pulsed dipole magnet.  The GPS time of the extraction magnet signal is recorded and defines the spill time~\cite{minosToF}.  

Each neutrino event was tagged with the local sidereal time (LST) of its spill -- the GPS spill time converted to sidereal time.  The local sidereal phase of an event is given by LST$\times (\omega_\oplus/2\pi)$ and has a range of 0-1.  Event times were not corrected for their time within a spill, an approximation that introduces no significant systematic error into the analysis.  

The events in each spill were placed into a single bin in a histogram spanning  0-1 in local sidereal phase (LSP).  The POT in the spill were binned into a second LSP histogram.  By dividing these two histograms, we get the number of $\nu_\mu$ events/POT as a function of LSP. This final histogram gives the normalized neutrino event rate in which we search for sidereal variations.

\begin{figure}
\centerline{\includegraphics[width=3.75in]{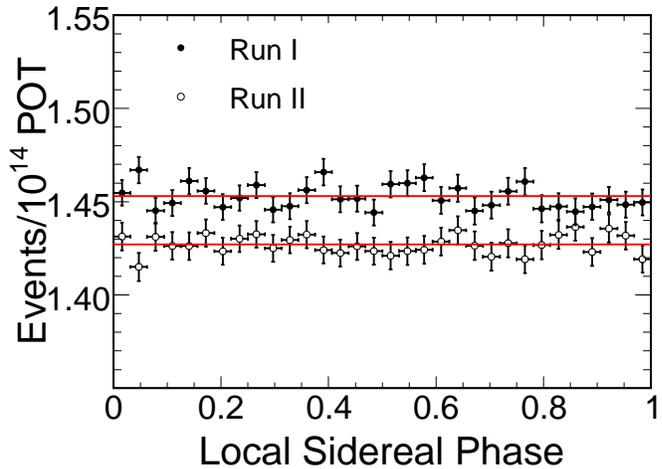}}
%\centerline{\epsfig{file=data_rate_power.eps,width=3.75in}}
\caption{\label{fig:data_rate}  The local sidereal phase histograms for Run~I and Run~II.  Superposed are fits to a constant sidereal rate.}
\end{figure}

We used 32 bins for the LSP histograms.  This binning was chosen to search for sidereal variations with a Fast Fourier Transform (FFT)~\cite{numrec} and the algorithm works most efficiently for $2^{\mathcal N}$ bins.   Since eq.(\ref{eq:osc}) puts power into harmonic frequencies associated with Fourier terms  to $n \omega_\oplus$, and for this analysis $n=$ 1-4, we chose ${\mathcal N} = 5$ as the minimal binning that retains these harmonic terms.   Each bin spans 0.031 in LSP or 45 min in sidereal time.  The histograms of the $\nu_\mu$ events/POT as a function of LSP for Run I and Run II are given in Fig.~\ref{fig:data_rate}.  The differences in the average event rates are due primarily to different relative positions of the target and magnetic focusing horns for the two runs.

We performed an FFT analysis on the Run I and Run II sidereal phase histograms in Fig.~\ref{fig:data_rate} and we computed the weighted mean of the powers returned for the even ($\cos{}$) and  odd ($\sin{}$)  powers for harmonic frequencies out to $  4 \omega_\oplus T_\oplus$.  The weighting factors were the mean event rates for each run.  The resulting mean powers, ${\bar p}(FFT)$, are listed in Table~\ref{table:sidPower}.   

\begin{table}
\caption{\label{table:sidPower} Weighted mean of Run~I and Run~II FFT powers in first four even/odd harmonic coefficients; ${\mathcal P}_F$ is the probability that the mean power is a noise fluctuation.}
\begin{tabular}{|ccc|ccc|}

\hline
 $\cos{()}$ & ~$\bar p$(FFT) ~& ~${\mathcal P}_F$~ & $\sin{()}$ &~$\bar p$(FFT) ~& ~${\mathcal P}_F$~ \\  \hline 

($\omega_\oplus T_\oplus$) &  -0.002   & 0.91 & ($\omega_\oplus T_\oplus$) &  0.024 & 0.18\\ 

($2 \omega_\oplus T_\oplus$)  & 0.011 & 0.54  & ($2 \omega_\oplus T_\oplus$) & 0.011  & 0.54 \\

($3 \omega_\oplus T_\oplus$) & -0.006 &  0.74  & ($3 \omega_\oplus T_\oplus$)  & -0.004 & 0.83 \\

($4 \omega_\oplus T_\oplus$)  & -0.016 & 0.37 & ($4 \omega_\oplus T_\oplus$)  & 0.023 & 0.20\\
\hline

\end{tabular}

\end{table}

These results were tested for several possible systematic effects.  We found that systematic increases or decreases in the event rate of 5\% in 6 months do not affect these results.  We also searched for systematic changes in the rates from day to night and found no variations $>$ 0.1\%.  In addition, we searched and found no sidereal modulation in the CC/NC ratio for these data.  This test shows that there are no systematic effects associated with neutrino production in the beam that affect this sidereal analysis.

We constructed 1,000 simulated experiments for both runs without a sidereal signal to test the significance of the powers given in Table~\ref{table:sidPower}.  We used the data themselves to construct these experiments.  We first generated 1,000 sets of sidereal phases for Run~I and Run~II, with each set having the same number of entries as spills in the run.  The phases were drawn randomly from the sidereal phase distribution constructed from the start times for each spill in the Run~I or Run~II data set.   We then put the events for each spill in the run into 1,000 separate histograms according to the scrambled sidereal phase assigned.  The number of POT for that spill was entered into a second set of 1,000 histograms according to the same set of sidereal phases.  The division of each event histogram by its corresponding POT histogram results in the simulated experiments -- histograms of  the number of $\nu_\mu$ events/POT as a function of LSP without a signal.  

\begin{figure}
\centerline{\includegraphics[width=3.75in]{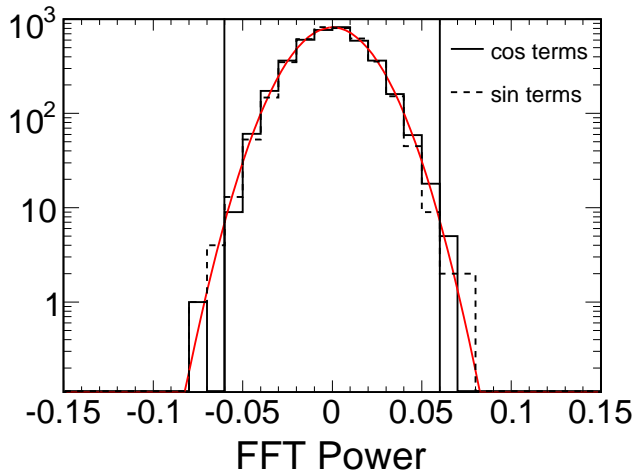}}
\caption{\label{fig:mc_power}  The distributions for the even ($\cos{}$) and the odd ($\sin{}$) mean powers for harmonic frequencies to $  4 \omega_\oplus T_\oplus$ from the FFT analysis of 1,000 simulated experiments in Run~I and Run~II. Superposed on these distributions is a Gaussian fit of width $\sigma = 1.8 \times 10^{-2}$.  This fit was obtained independently for both distributions.  Values outside of the vertical lines are more than $3\sigma$ from the mean.
}
\end{figure}

We performed the same FFT analysis on each of the 1,000 Run~I and 1,000 Run~II simulated experiments as was done with the sidereal phase histograms in Fig.~\ref{fig:data_rate}.  The powers returned by these FFTs give the fluctuation spectrum expected from sidereal phase histograms in which there is no sidereal signal.  As for the data, we computed the weighted mean power for each harmonic in a pair of simulated Run~I  and Run~II experiments.  The distributions for the even ($\cos{}$) and the odd ($\sin{}$) mean powers for harmonic frequencies out to $  4 \omega_\oplus T_\oplus$ are shown superposed in Fig.~\ref{fig:mc_power}.   Clearly these even and odd distributions are nearly identical.  In addition, a Gaussian of width $\sigma = 1.8 \times 10^{-2}$ has been superposed onto these two distributions in Fig.~\ref{fig:mc_power}.  This fit was obtained independently for both distributions.  We use this Gaussian to estimate the probability that the powers returned by the FFT analysis of the sidereal phase histograms in Fig.~\ref{fig:data_rate} are due to statistical fluctuations.  

Table~\ref{table:sidPower} gives the probability, ${\mathcal P}_F$, that the mean power represents a noise fluctuation.  It was calculated as the probability of drawing a value of the weighted mean power for the two data sets at least as large as found from the parent Gaussian distribution in Fig.~\ref{fig:mc_power}.  Since the largest fluctuation in the FFT power in the Fig.~\ref{fig:data_rate} histogram is $1.32 \sigma$ we conclude that no term reaches the level of a $3 \sigma$  detection.  We have determined that these results are insensitive to the exact choice of the zero point of sidereal phase.  This model-independent result implies that there is no significant change in normalized neutrino event rate that depends on the direction of the neutrino beam in a sun-centered inertial frame.  In the context of the SME, this result is inconsistent with the detection of LV.

In the absence of a sidereal signal, we can establish upper limits on the SME coefficients $(a_L)^\mu$ and $(c_L)^{\mu \nu}$ that describe LV~\cite{KM} using the standard MINOS Monte Carlo simulation.  The simulation includes weighting to account for hadron production off the NuMI target~\cite{Adamson:2007gu}.  In this simulation, events are generated by modeling the NuMI beam line, including the hadron production by the 120 GeV$/c$ protons on target, the propagation of the hadrons through the focusing elements and 675 m decay pipe to the beam absorber, and the calculation of the probability that any neutrinos generated traverse the ND.  The ND neutrino event simulation takes the neutrinos from the NuMI simulation, along with an energy determined by decay kinematics, and uses this information as input into the simulation of the ND.  With the known $L$ and $E_\nu$ for the simulated neutrino events, as well as the beam direction, we can inject a Lorentz-violating signal into eq.(\ref{eq:osc}).  The construction of MC-generated sidereal phase histrograms is described elsewhere~\cite{cpt07}. 

The limits on the LV coefficients $(a_L)^\mu$ and $(c_L)^{\mu \nu}$ were determined from a set of 200 simulated experiments.  First we set all but one LV coefficient to zero.  We next weighted the simulated neutrino events in each histogram by its survival probability computed according to eq.(\ref{eq:osc}), assuming the LV coefficient is small.  We then increased the magnitude of the nonzero coefficient until one of the FFT powers in the simulated phase histogram was $3\sigma$ away from the mean of the distribution in Fig.~\ref{fig:mc_power}.  An average of the 200 determinations of each SME coefficient, scaled in terms of the suppression factor  $m_W/m_p \sim 10^{-17}$, is given in Table~\ref{table:lorentzCoeff}.  This procedure, by which we vary one parameter at a time to determine the limits, could miss fortuitous cancellations of SME coefficients thereby masking a signal.  However, we consider such cancellations in Nature to be highly unlikely.  
 
\begin{table}
\caption{\label{table:lorentzCoeff}  Limits to SME coefficients for $\nu_\mu \rightarrow \nu_x$ in terms of the suppression factor  $m_W/m_P \sim 10^{-17}$; $a_L$ have units of (GeV) and $c_L$ are unitless.} 

\begin{tabular}{|lc|lc|} 

\hline \hline

 & ~$\times 10^{-17}$ ~&  &~$\times 10^{-17}$ ~\\ \hline 

$a_L^{X}$ &  $3.0 \times 10^{-3}$ & $a_L^Y$ & $3.0 \times 10^{-3}$ \\  

$c_L^{TX}$ &  $0.9 \times 10^{-5}$ & $c_L^{TY}$ &  $0.9 \times 10^{-5}$ \\

$c_L^{XX}$ &  $5.6 \times 10^{-4}$ & $c_L^{YY}$ &  $5.5 \times 10^{-4}$ \\

$c_L^{XY}$ & $2.7 \times 10^{-4}$ & $c_L^{YZ}$ &  $1.2\times 10^{-4}$ \\

$c_L^{XZ}$ &  $1.3 \times 10^{-4}$ & ~~--~~ &~~~ -- ~~~\\

\hline \hline

\end{tabular}

\end{table}

In summary, we find no significant evidence for sidereal variations in the MINOS ND neutrino data.  When framed in the SME theory~\cite{KM}, this result leads to the conclusion that we have detected no evidence for the violation of Lorentz and CPT invariance.  Based on these results, we computed limits on the LV SME coefficients and find that their magnitude is $<$ 1\% of the suppression factor $m_W/m_P \sim 10^{-17}$.  For the $a_L$-type SME coefficients, the MINOS limits are a factor of 3 lower than those reported by LSND~\cite{LSND}; for the $c_L$-type SME coefficients, the MINOS limits are at least 4 orders of magnitude lower than LSND's.
  
We gratefully acknowledge the many valuable conversations with Alan Kosteleck\'y during the course of this work.  This work was supported by the US DOE, the UK STFC, the US NSF, the State and University of Minnesota, the University of Athens, Greece, and Brazil's FAPESP and CNPq.  We are grateful to the Minnesota Department of Natural Resources, the crew of the Soudan Underground Laboratory, and the staff of Fermilab for their contribution to this effort.
 
\bibliography{lorentzPRL}

\begin{thebibliography}{16}
\expandafter\ifx\csname natexlab\endcsname\relax\def\natexlab#1{#1}\fi
\expandafter\ifx\csname bibnamefont\endcsname\relax
  \def\bibnamefont#1{#1}\fi
\expandafter\ifx\csname bibfnamefont\endcsname\relax
  \def\bibfnamefont#1{#1}\fi
\expandafter\ifx\csname citenamefont\endcsname\relax
  \def\citenamefont#1{#1}\fi
\expandafter\ifx\csname url\endcsname\relax
  \def\url#1{\texttt{#1}}\fi
\expandafter\ifx\csname urlprefix\endcsname\relax\def\urlprefix{URL }\fi
\providecommand{\bibinfo}[2]{#2}
\providecommand{\eprint}[2][]{\url{#2}}

\bibitem[{oth()}]{others}
\bibinfo{note}{G. Amelino-Camelia et al., AIP Conf. Proc. {\bf 758}, 30 (2005),
  gr-qc/0501053; R. Bluhm, Lec. Notes Phys. {\bf 702}, 191 (2006),
  hep-ph/0506054}.

\bibitem[{CKc()}]{CKcomb}
\bibinfo{note}{D. Colladay and V. A. Kosteleck\'y, Phys. Rev. D {\bf 55}, 6760
  (1997); D. Colladay and V. A. Kosteleck\'y, Phys. Rev. D {\bf 58}, 116002
  (1998); V. A. Kosteleck\'y, Phys. Rev. D {\bf 69}, 105009 (2004)}.

\bibitem[{\citenamefont{Kosteleck\'y and Mewes}(2004{\natexlab{a}})}]{KM2}
\bibinfo{author}{\bibfnamefont{V.~A.} \bibnamefont{Kosteleck\'y}}
  \bibnamefont{and} \bibinfo{author}{\bibfnamefont{M.}~\bibnamefont{Mewes}},
  \bibinfo{journal}{Phys. Rev. D} \textbf{\bibinfo{volume}{69}},
  \bibinfo{pages}{0160005} (\bibinfo{year}{2004}{\natexlab{a}}).

\bibitem[{lit()}]{lit}
\bibinfo{note}{S. Reinhardt et al., Nature Physics {\bf 3}, 861 (2007); H.
  Mueller et al., Phys. Rev. Lett., {\bf 99}, 050401 (2007); V. W. Hughes et
  al., Phys. Rev. Lett., {\bf 87}, 111804 (2001); Y. B. Hsiung et al., Nucl.
  Phys. Proc. Suppl, {\bf 86}, 312 (2000); J. Link et al., Phys. Lett. B, {\bf
  556}, 7, (2003); B. Aubert et al., Phys. Rev. D, {\bf 70}, 012007 (2004);
  J.B.R. Battat et al., Phys. Rev. Lett., {\bf 99}, 241103 (2007); P. Wolf et
  al., Phys. Rev. Lett., {\bf 96}, 060801 (2006); H. Demelt et al., Phys. Rev.
  Lett, {\bf 83}, 4694 (1999); B. Heckel et al., Phys. Rev. Lett., {\bf 97},
  021603 (2006); M.D. Messier, in {\it Proceedings of the Third Meeting on CPT
  and Lorentz Symmetry}, edited by V.A. Kosteleck\'y (World Scientific, 2005),
  p.84.}

\bibitem[{\citenamefont{Kosteleck\'y and Mewes}(2004{\natexlab{b}})}]{KM}
\bibinfo{author}{\bibfnamefont{V.~A.} \bibnamefont{Kosteleck\'y}}
  \bibnamefont{and} \bibinfo{author}{\bibfnamefont{M.}~\bibnamefont{Mewes}},
  \bibinfo{journal}{Phys. Rev. D} \textbf{\bibinfo{volume}{70}},
  \bibinfo{pages}{076002} (\bibinfo{year}{2004}{\natexlab{b}}).

\bibitem[{\citenamefont{Adamson et~al.}(2008{\natexlab{a}})}]{Adamson:2007gu}
\bibinfo{author}{\bibfnamefont{P.}~\bibnamefont{Adamson}} \bibnamefont{et~al.},
  \bibinfo{journal}{Phys. Rev. D} \textbf{\bibinfo{volume}{77}},
  \bibinfo{pages}{072002} (\bibinfo{year}{2008}{\natexlab{a}}).

\bibitem[{\citenamefont{Auerbach et~al.}(2005)}]{LSND}
\bibinfo{author}{\bibfnamefont{L.~B.} \bibnamefont{Auerbach}}
  \bibnamefont{et~al.}, \bibinfo{journal}{Phys. Rev. D}
  \textbf{\bibinfo{volume}{72}}, \bibinfo{pages}{076004}
  (\bibinfo{year}{2005}).

\bibitem[{\citenamefont{Michael et~al.}(2006)}]{minos2}
\bibinfo{author}{\bibfnamefont{D.~G.} \bibnamefont{Michael}}
  \bibnamefont{et~al.}, \bibinfo{journal}{Phys. Rev. Lett.}
  \textbf{\bibinfo{volume}{97}}, \bibinfo{pages}{191801}
  (\bibinfo{year}{2006}).

\bibitem[{\citenamefont{Greenberg}(2002)}]{G1}
\bibinfo{author}{\bibfnamefont{O.~W.} \bibnamefont{Greenberg}},
  \bibinfo{journal}{Phys. Rev. Lett.} \textbf{\bibinfo{volume}{89}},
  \bibinfo{pages}{231602} (\bibinfo{year}{2002}).

\bibitem[{\citenamefont{Nunokawa et~al.}(2005)\citenamefont{Nunokawa, Parke,
  and Zukanovich~Funchal}}]{Nunokawa:2005nx}
\bibinfo{author}{\bibfnamefont{H.}~\bibnamefont{Nunokawa}},
  \bibinfo{author}{\bibfnamefont{S.~J.} \bibnamefont{Parke}}, \bibnamefont{and}
  \bibinfo{author}{\bibfnamefont{R.}~\bibnamefont{Zukanovich~Funchal}},
  \bibinfo{journal}{Phys. Rev. D} \textbf{\bibinfo{volume}{72}},
  \bibinfo{pages}{013009} (\bibinfo{year}{2005}), \eprint{hep-ph/0503283}.

\bibitem[{\citenamefont{Michael et~al.}(2008)}]{minosD}
\bibinfo{author}{\bibfnamefont{D.~G.} \bibnamefont{Michael}}
  \bibnamefont{et~al.} (\bibinfo{year}{2008}), \eprint{arXiv:0805.3170
  [physics.ins-det]}.

\bibitem[{\citenamefont{Tagg et~al.}(2005)}]{Tagg:2004bu}
\bibinfo{author}{\bibfnamefont{N.}~\bibnamefont{Tagg}} \bibnamefont{et~al.},
  \bibinfo{journal}{Nucl. Instrum. Meth.} \textbf{\bibinfo{volume}{A539}},
  \bibinfo{pages}{668} (\bibinfo{year}{2005}), \eprint{physics/0408055}.

\bibitem[{\citenamefont{Adamson et~al.}(2008{\natexlab{b}})}]{minosNC}
\bibinfo{author}{\bibfnamefont{P.}~\bibnamefont{Adamson}} \bibnamefont{et~al.}
  (\bibinfo{year}{2008}{\natexlab{b}}), \bibinfo{note}{in preparation}.

\bibitem[{\citenamefont{Adamson et~al.}(2007)}]{minosToF}
\bibinfo{author}{\bibfnamefont{P.}~\bibnamefont{Adamson}} \bibnamefont{et~al.},
  \bibinfo{journal}{Phys. Rev. D} \textbf{\bibinfo{volume}{76}},
  \bibinfo{pages}{072005} (\bibinfo{year}{2007}).

\bibitem[{\citenamefont{Press et~al.}(1999)\citenamefont{Press, Flannery,
  Teukolsky, and Vetterling}}]{numrec}
\bibinfo{author}{\bibfnamefont{W.~H.} \bibnamefont{Press}},
  \bibinfo{author}{\bibfnamefont{B.}~\bibnamefont{Flannery}},
  \bibinfo{author}{\bibfnamefont{S.}~\bibnamefont{Teukolsky}},
  \bibnamefont{and}
  \bibinfo{author}{\bibfnamefont{W.}~\bibnamefont{Vetterling}},
  \emph{\bibinfo{title}{Numerical Recipes in C}} (\bibinfo{publisher}{Cambridge
  University Press}, \bibinfo{year}{1999}).

\bibitem[{\citenamefont{Rebel and Mufson}(2008)}]{cpt07}
\bibinfo{author}{\bibfnamefont{B.~J.} \bibnamefont{Rebel}} \bibnamefont{and}
  \bibinfo{author}{\bibfnamefont{S.~L.} \bibnamefont{Mufson}}, in
  \emph{\bibinfo{booktitle}{Proceedings of the Fourth Meeting on CPT and
  Lorentz Symmetry}}, edited by \bibinfo{editor}{\bibfnamefont{V.~A.}
  \bibnamefont{Kosteleck\'y}} (\bibinfo{publisher}{World Scientific},
  \bibinfo{year}{2008}), \eprint{arXiv:0802.3785 [hep-ph]}.

\end{thebibliography}

\end{document}